\documentclass{osa-article-1}

\journal{osac}

\articletype{Research Article}

\begin{document}

\title{Quantum key distribution with multiphoton pulses: An advantage}

\author{Ayan Biswas\authormark{1,2}, Anindya Banerji\authormark{1,*}, Nijil Lal\authormark{1}, Pooja Chandravanshi\authormark{1}, Rupesh Kumar\authormark{3} and Ravindra P. Singh\authormark{1,4}}

\address{\authormark{1}Quantum Science and Technology Laboratory, Physical Research Laboratory, Ahmedabad 380009, India\\
\authormark{2}Indian Institute of Technology, Gandhinagar 382355, India\\
\authormark{3}Quantum Communications Hub and York Centre for Quantum Technologies, Department of Physics, University of York, York, YO10 5DD, UK\\
\authormark{4}rpsingh@prl.res.in}

\email{\authormark{*}abanerji09@gmail.com}

%%%%%%%%%%%%%%%%%%% abstract %%%%%%%%%%%%%%%%

\begin{abstract}
 In this article, we introduce a quantum key distribution protocol for the line of sight channels based on coincidence measurements. We present a proof-of-concept implementation of our protocol. We show that using coincidence measurements to monitor multi-photon pulses results in a higher secure key rate over longer distances for such channels. This key rate is higher than popular implementations of quantum key distribution protocol based on BB84, for example, the GLLP analysis [Quant. Info. Comput. \textbf{4}, 325 (2004)]. In the experiment, we could generate around $74 \%$ more key bits per signal pulse as compared to the GLLP analysis of BB84 protocol with similar parameters and equal value of mean photon number.
\end{abstract}

%%%%%%%%%%%%%%%%%%%%%%%%%%  body  %%%%%%%%%%%%%%%%%%%%%%%%%%
\section{Introduction}
Quantum key distribution \cite{RevModPhys.74.145,RevModPhys.81.1301,arxiv:1906.01645} is perhaps the most remarkable application of quantum theory. It exploits the principles of quantum mechanics to enable two distant parties to share a secret random key. Once the key has been established, the two can exchange encrypted messages using private key cryptographic methods. BB84 \cite{BENNETT20147} was the first QKD protocol, based on two basis of orthogonal states, experimentally realised \cite{J.Cryptology.5.3} followed by proposals with two nonorthogonal states \cite{PhysRevLett.68.3121} and the entanglement based protocol \cite{PhysRevLett.67.661}. BB84 is proven to be unconditionally secure, based solely on the validity of the laws of quantum mechanics \cite{J.Acm.48, PhysRevLett.85.441, Mayers2002}. It was later pointed out that imperfection in practical implementations seriously undermine the security of the QKD protocols \cite{PhysRevLett.85.1330}. This led to proposals for various types of attacks exploiting the imperfections in the components of the QKD system \cite{JModOpt.48.2023, NewJPhys.12.113026, JModOpt.48.2009, NewJPhys.11}. One of them was the lack of ideal single photon sources. This led to the use of weak coherent pulses in which the number of photons in each pulse is governed by a Poissonian distribution. This leads to non-zero probability of pulses containing more than one photon. An eavesdropper can exploit this major vulnerability to extract information about the key during the transmission stage by using a photon number splitting attack \cite{PhysRevLett.85.1330}. This resulted in several innovative protocols \cite{PhysRevLett.94.230504, PhysRevLett.92.057901, PhysRevA.73.010302, PhysRevLett.94.230503, PhysRevLett.90.057902,PhysRevLett.108.130503,PhysRevLett.113.140501, PhysRevA.88.032305,Nature.557.400,Optica.4.1006} and proof of security with practical implementations \cite{PhysRevA.61.052304,Gottesman:2004:SQK:2011586.2011587,Inamori2007}. Notable among the proposed protocols was the decoy state protocol \cite{PhysRevLett.94.230504, PhysRevLett.91.057901} for its efficient mitigation of the photon number splitting attack. On the other hand, entanglement based protocols \cite{PhysRevLett.108.130503,PhysRevLett.113.140501} suffered from very low key rates and problem of distributing entanglement over long distances reliably with high fidelity. As a result, the decoy state method emerged as the preferred method for long distance quantum key distribution \cite{PhysRevLett.121.190502,Nature.549,Dixon:15} with a key rate that was substantially higher than the key rate for implementations with imperfect devices \cite{Gottesman:2004:SQK:2011586.2011587}. In this method, the sender, Alice, prepares a set of decoy pulses with varying intensities in addition to the standard BB84 states. The decoy pulses are inserted randomly within the actual signal pulse train unknown to the receiver, Bob, as well as any potential eavesdropper, Eve. Without any prior knowledge regarding the position of the decoy pulses, there is an equal probability of Eve attacking both the decoy as well as the BB84 signal pulses. By monitoring the quantum bit error rate (QBER) of the decoy pulses, Alice and Bob can reliably estimate a lower bound for the secret key rate. But the improved performance comes at a cost. Implementation of the decoy state protocol requires multiple intensities of the weak coherent pulses, its calibration and increased complexities in hardware and processing.\\
The major contribution of this article is to demonstrate that, an increased key rate can be achieved without using decoy pulses when communicating parties are in direct line of sight (LOS) channel which can be monitored by other methods, for example, using Lidars \cite{QCrypt.2017, QCrypt.2019}. LOS channel are most commonly used in terrestrial communication between two towers in the same city or different cities and also in high altitudes \cite{chu2020feasibility , SNureth} . For small distance communication direct LOS channel can be realized through the drones \cite{SIsaac}. The protocol utilises the inherent randomness in the number of photons per pulse of the source itself. Even if eavesdropper is injecting photons directed towards Bob’s receiver, it would result in increasing the two-photon and three-photon error in the coincidence detection. The presence of multi-photon pulses sent by Alice are tracked by coincidence detection at Bob's end and secure key is extracted using some of the multi-photon pulses too. The difference in the number of actual recorded coincidences and expected number of coincidences for a given value of mean photon number for a given channel plays an important factor in this case. If the ratio of this difference with the actual number of coincidence increases above a threshold value, security is compromised and the protocol is aborted. Otherwise they form the key from the single as well as some of the multi-photon pulses followed by standard error correction and privacy amplification methods. We also use an additional figure of merit, the ratio of coincidences to singles to further monitor the security. Since we use coincidence measurements as a major tool, we call this the Coincidence Detection (CD) protocol.

\section{Mathematical Background}

In this section, we will provide the mathematical derivation of the key rate for our protocol. In what follows, we will compare the key rates of our proposed protocol with the GLLP analysis \cite{Gottesman:2004:SQK:2011586.2011587} since both these protocols make use of a single value of mean photon number. This makes our approach distinct from the decoy state method which uses more than one value of mean photon number. But before proceeding with the derivation, let us first briefly outline the protocol as follows: Alice sends weak coherent pulses to Bob prepared in the standard way for polarization based implementations of BB84. Since the number of photons in each pulse is governed by poissonian statistics, some of the pulses might contain more than one photon. Neither Alice nor Bob has any control over the occurrence of these pulses. Instead of looking at this inherent randomness in the photon number distribution as a drawback, we use it to our advantage. Bob, while recording the measurement results, also records all the 2 and 3-fold coincidence events. The coincidence window is set according to the pulse width of the signal pulses. The total number of coincidences are matched with the expected number of coincidences which are calculated from the value of $\mu$. It was already shown in \cite{JModOpt.48.2009} that the coincidences arising from multiphoton pulses can be tracked to ensure no information is leaked to Eve. Any change in the number of 2 and 3-fold coincidences than the expected value for a specific channel will reveal the presence of eavesdropper in the system assuming that Eve is randomly attacking the pulse (no collective and coherent attack). To estimate the number of 2 and 3-fold coincidence events, it is essential to consider how the pulses split at a balanced beam splitter. For \emph{n} photon input state, the photons are distributed between the reflected and transmitted ports as
\begin{equation}
    \vert n\rangle \rightarrow \sum_{k=0}^n C_k^n \vert n-k \rangle_R\vert k \rangle_T,
\end{equation}
\noindent where $R$($T$) corresponds to the reflected (transmitted) port. $\vert C_k^n\vert^2$ is the probability of getting \emph{n-k} (\emph{k}) photons in the reflected (transmitted) port. The possible cases for 2 and 3 photon pulses are given below in the tables \ref{tab:2photons} and \ref{tab:3photons} respectively. We will take the coincidences arising out of this splitting of pulses into our consideration when deriving the final key rate.\\
\begin{table}[h!]
\centering
\caption{Splitting of a two-photon pulse at a beam splitter.}
\begin{tabular}{cccc}
\hline
\begin{tabular}[c]{@{}c@{}}Possible\\ Cases\end{tabular} 
& \begin{tabular}[c]{@{}c@{}}Number of\\ Photons at\\ Transmitted Port\end{tabular} 
& \begin{tabular}[c]{@{}c@{}}Number of\\ Photons at\\ Reflected Port\end{tabular} 
& Probability \\ 
\hline
1   & 2      & 0 & 1/4    \\
2   & 0      & 2 & 1/4     \\
3   & 1      & 1 & 1/2   \\
\hline     
\end{tabular}
\label{tab:2photons}
\end{table}
\begin{table}[h!]
\centering
\caption{Splitting of a three-photon pulse at a beam splitter.}
\begin{tabular}{cccc}
\hline
\begin{tabular}[c]{@{}c@{}}Possible\\ Cases\end{tabular} 
& \begin{tabular}[c]{@{}c@{}}Number of\\ Photons at\\ Transmitted Port\end{tabular} 
& \begin{tabular}[c]{@{}c@{}}Number of\\ Photons at\\ Reflected Port\end{tabular} 
& Probability \\ 
\hline
1   & 3      & 0 & 1/8    \\
2   & 0      & 3 & 1/8     \\
3   & 1      & 2 & 3/8      \\
4   & 2      & 1 & 3/8\\
\hline
\end{tabular}
\label{tab:3photons}
\end{table}

In order to derive the key rate, we follow the treatment of \cite{PhysRevLett.94.230504}. We denote phase randomized signal state of the weak coherent pulses as mixture of coherent states
\begin{equation}
   \rho = \frac{1}{2 \pi}\int_{0}^{2 \pi} \vert \sqrt{\mu}e^{i \theta}\rangle \langle \sqrt{\mu}e^{i \theta}\vert d \theta.
\end{equation}
Here, $\mu$ stands for average number of photons per pulse and the signal is assumed to be randomised over all $\theta$. The probability $P(n)$ of each pulse carrying \emph{n} photons is derived from the Poissonian distribution as $P(n)=e^{-\mu}\mu^{n}/n!$. Progressing onwards, the gain $Q_{\mu}$ of each pulse is defined as
\begin{equation}
\label{gain}
    \begin{split}
        Q_{\mu}=& Y_{0}e^{-\mu}+Y_{1}e^{-\mu}\mu+Y_{2}e^{-\mu}(\mu^{2}/2!)+...
        +...+Y_{n}e^{-\mu}(\mu^{n}/n!),
    \end{split}
\end{equation}
where $Y_n$ is the conditional probability that Bob detects an ``n photon" signal state given that Alice has sent  an ``n photon" state. Then, $Q_n$ becomes the joint probability of Bob detecting ``n photon" signal and Alice sending the same ``n photon" signal state. For realistic cases, in the absence of an eavesdropper, the term $Y_0$ gives the background rate of the system including detector dark counts, $p_{dark}$. For $n \geq 1$ , yield $Y_n$ consists of two terms, the detection of signal photons travelling through the channel and the background rate. Assuming that the background rate and the signal events are independent, the expression of $Y_n$ is seen to be dependent on the channel \cite{PhysRevLett.94.230504} and approximated to
\begin{equation}
\label{yield}
   Y_n \approx [\eta_n + p_{dark}]/2.
\end{equation}
The transmission efficiency $\eta_n$ of the channel is related to the number of photons as
\begin{equation}
    \eta_n=1-(1- \eta)^n,
\end{equation}
\noindent where $\eta$ is the overall channel transmissivity. Now, the quantum bit error rate (QBER) corresponding to each signal state, $E_{\mu}$, is defined as
\begin{equation}
\label{QBER}
     E_{\mu}Q_{\mu}= \sum_{n=0}^{\infty}Q_n E_n,
\end{equation}
\noindent where $E_n$ is the error corresponding to the signal containing \emph{n} photons. Even in the absence of any signal pulse, Bob might record a detection due to background photons or dark current of the detector. This error results in $E_0$ and is equal to 1/4 since all four detectors have equal probability of registering a dark count. If the signal has $n \geq 1$ photons, then the error $E_n$ is given by 
\begin{equation}
    E_{n}=\big( \eta_n \frac{E_{detector}}{2}+(1-\eta_n)\frac{p_{dark}}{4} \big)/Y_n,
\end{equation}

\noindent where $E_{detector}$ is independent of $n$ and the values of $E_n$ and $Y_n$ can be experimentally derived from the measured values of $Q_{\mu}$ and $E_{\mu}$. Major change in these values for a specific channel will reveal the presence of eavesdropper.\\ 
Having defined all the necessary terms and variables, let us briefly look at how the equations governing the secret key rate evolves. It was shown in \cite{PhysRevLett.85.441} that secret key rate in an ideal implementation scenario with a perfect single photon source and perfect detectors has the form
\begin{equation}
\label{keyrate_basic}
    R \geq [1-2H_2(E_b)],
\end{equation}
\noindent where $H_2$ is the binary Shannon entropy defined as $H_2(x)=-x log_{2}x- (1-x) log_{2}(1-x) $ and $E_b$ is the QBER. This formula was later modified by  \cite{Gottesman:2004:SQK:2011586.2011587} for a more realisitic implementation with weak coherent pulses as
\begin{equation}
\label{keyrate_GLLP}
     R \geq q Q_{\mu}\Big\{-f(E_{\mu})H_2(E_{\mu})+\frac{Q_1}{Q_{\mu}} \Big[1-H_2\big(\frac{Q_{\mu}E_{\mu}}{Q_1}\big)\Big] \Big\},
\end{equation}
\noindent where \emph{q} is an implementation dependent factor. In case of passive random basis selector, like balanced beam splitter, $q=1/2$. $f(E_{\mu})$ is the error correcting code efficiency. A severe shortcoming of the above approach was in estimating the maximal value of $\mu$. In order to minimise the number of pulses with 2 or above photons, $\mu$ had to be kept sufficiently small. This reduced the number of single photon pules thereby greatly limiting the secret key rate. At the same time, the protocol was vulnerable \textcolor{black}{to} PNS attacks since the absence of multiphoton pulses could not be ensured. In the decoy state protocol \cite{PhysRevLett.94.230504}, this was taken care of and the secret key rate was modified to
\begin{equation}
\label{keyrate_Decoy}
    R \geq q \{-Q_{\mu}f(E_{\mu})H_2(E_{\mu})+Q_{1}[1-H_2(E_1)] \}.
\end{equation}
\subsection{Key Rate Estimation for Coincidence Detection Method}

\textcolor{black}{It is seen in Eq \ref{keyrate_Decoy} that only single photons are contributing to the key}. Now, instead of discarding all the multiphoton pulses, we systematically include a fraction of all such pulses in the final secret key rate as
\begin{equation}
\label{keyrate_CD}
    \begin{split}
        R_{CD} \geq & \{-qQ_{\mu}f(E_{\mu})H_2(E_{\mu})+C_1 Q_{1}[1-H_2(E_1)]\\
        & +C_2 Q_{2}[1-H_2(E_2)]+C_3 Q_{3}[1-H_2(E_3)]\},
    \end{split}
\end{equation}
where $C_n$'s are the coefficients of the contributing single, double and triple photons pulses with the implementation dependent factor \emph{q} absorbed into them. This is the secret key rate of the CD protocol. In order to derive these coefficients, consider the following: a single photon pulse can only end in the correct basis with probability 1/2 in case of passive basis selector like a balanced beam spliter \textcolor{black}{for which $q = 1/2$. This leads to} to $C_1= 1/2$. A two-photon pulse will give rise to three cases as in Table \ref{tab:2photons} of which case 3 and only one of case 1 \textcolor{black}{or} case 2 will contribute to the key. So, $C_2=1/2+1/4=3/4$. Similarly, from Table \ref{tab:3photons} we obtain $C_3=3/8+3/8+1/8=7/8$. In this case, both cases 3 and 4 will contribute to the key since in both cases at least one photon will be detected in the correct basis. \textcolor{black}{Please note that the probabilities in Tables \ref{tab:2photons} and \ref{tab:3photons} are calculated for a balanced beam splitter. So the factor of $q = 1/2$ is already accounted for while calculating the probabilities justifying the absorption of $q$ into $C_n$.} Substituting these values in Eq. \ref{keyrate_CD} we arrive at the final form of the secret key rate. The final secret key rate is as follows
\begin{equation}
\label{keyrate_final}
    \begin{split}
        R_{CD} \geq & \{ -\frac{1}{2}Q_{\mu}f(E_{\mu})H_2(E_{\mu})+\frac{1}{2} Q_{1}[1-H_2(E_1)]\\
        &+\frac{3}{4} Q_{2}[1-H_2(E_2)]+\frac{7}{8} Q_{3}[1-H_2(E_3)]\}.
    \end{split}
\end{equation}
It is evident that some of the pulses with multiple photons also contribute, leading to a higher secret key rate. This protocol, a modification of BB84 protocol, works best with four SPCMs (Single Photon Counting Modlues) as more number of multiphoton pulses can be tracked and the keys can be extracted from them. For two detector system Eq. 12 will be modified by omitting the last term as only two fold coincidences will be observed. For single detector setup only the first and second term will remain in Eq. 12.
\subsection{Security against Eavesdropper}
The standard security analysis of a QKD protocol involves calculating the difference in mutual information between the communicating parties and the eavesdropper. For direct reconciliation (DR) the difference in mutual infromation between Alice-Bob and Alice-Eve while, it is Alice-Bob and Bob-Eve for reverse reconciliation (RR). If the mutual information between Alice-Bob exceeds that between Alice-Eve (DR) or Bob-Eve (RR), a secure key can be extracted and the channel is deemed secure. An additional parameter is the QBER. For BB84 based protocols using ideal source and detector, the QBER has an upper limit of 11$\%$ against collective attakcs \cite{arxiv:1906.01645,PhysRevLett.85.441}. After the protocol is executed, if the estimated QBER exceeds that limit, the channel is discarded and the protocol is repeated again. For all those attacks that affect the QBER, it serves as a powerful tool at the hands of the communicating parties. \\
The security for our protocol is derived from monitoring the QBER as well as the total number of coincidences for a \emph{pre-characterised channel} within acceptable statistical fluctuations due to device and channel limitations. This means, that the channel transmittance is known and is trusted. This is ensured by actively monitoring the channel during the characterisation process. The total number of coincidences expected are 
\begin{equation}
    \label{Eq.1:Actual_Coincidence}
    C = \frac{1}{2}Y_2P_2\left(\mu\right) + \frac{3}{4}Y_3P_3\left(\mu\right).
\end{equation}

\noindent $P_n\left(\mu\right)$ is the Poissonian probability of a pulse containing \emph{n} photons for a given $\mu$. Since the yield $Y_n$ depends on $\eta$, the coincidences depend on both $\mu$ and $\eta$. The fractions $\left(1/2\right)$ and $\left(3/4\right)$ arise due to the use of a balanced (50:50) beam splitter as the basis selector. The equation (\ref{Eq.1:Actual_Coincidence}) means that a two-photon pulse will produce a coincidence half of the times while a three-photon pulse will result in a coincidence 3 out of 4 times. Now, the yields are related as already seen in Eq \ref{yield}

\begin{equation}
    \label{Eq.2:Yields}
    Y_2 = 2\eta ; Y_3 = 3\eta  = \frac{3}{2}Y_2.
\end{equation}

\noindent Substituting these values and writing $P_3(\mu)$ in terms $P_2(\mu)$ in Eq. (\ref{Eq.1:Actual_Coincidence}), we can write the total number of coincidences as 
\begin{eqnarray}
    \label{Eq.4:Modified_Coincidence}
    C = \left(\frac{4 + 3\mu}{8}\right)Y_2P_2\left(\mu\right).
\end{eqnarray}

\noindent Now, we can define a figure of merit $\Xi = \Delta C/C$, which is the ratio of change in coincidences to the total number of coincidences. Under normal circumstances, the coincidences will vary due to statistical fluctuations. As the number of coincidences depend on $\eta$ and $\mu$, the statistical fluctuation $\Delta C_{stat}$ can be written as as

\begin{equation}
    \label{Eq.9:Statistical_fluctuation}
    \Delta C_{stat}= \left|\frac{\partial C}{\partial \eta}\right|\Delta \eta+\left|\frac{\partial C}{\partial \mu}\right| \Delta \mu.
\end{equation}

\noindent Both these terms can be rewritten in terms of the $Y_2$ and $P_2\left(\mu\right)$. This will help us to compare $\Delta C_{stat}$ with $C$. Making the necessary substitutions, we arrive at the form

\begin{equation}
    \label{Eq.12:Statistical_fluctuation}
    \Delta C_{stat} = \frac{8 + 5\mu -3\mu^2}{8\mu}Y_2P_2\left(\mu\right)\Delta \mu + \frac{4 + 3\mu}{8\eta}Y_2P_2\left(\mu\right)\Delta \eta.
\end{equation}

\noindent The factors $\Delta \mu$ and $\Delta \eta$ are implementation dependent factors. The fluctuation in $\mu$ can arise from imperfect attenuators while fluctuations in $\eta$ can arise due to atmospheric changes. Let us assume that $\mu$ varies by a factor of $\alpha$ over the duration for which the protocol is run i.e. $\Delta \mu = \alpha \mu$. For the same duration, let the transmissivity vary by a factor of $\beta$. So, $\Delta \eta = \beta \eta$. Under these conditions, Eq. (\ref{Eq.12:Statistical_fluctuation}) is given by

\begin{equation}
    \label{Eq.13:Stat_fluc_final}
    \Delta C_{stat} = \frac{\alpha\left(8 + 5\mu -3\mu^2\right) + \beta\left(4 + 3\mu\right)}{8}Y_2P_2\left(\mu\right).
\end{equation}

\noindent We can now define a bound for $\Xi$ that takes into account the statistical fluctuations. We denote it by $\Xi_{stat} = \Delta C_{stat}/C$, which has the form

\begin{equation}
    \label{Eq.14:FOM_stat}
    \Xi_{stat} = \frac{8\alpha + 4\beta + \left(5\alpha + 3\beta\right)\mu - 3\alpha\mu^2}{4 + 3\mu}.
\end{equation}

\noindent Assuming that the experimental conditions do not change much during the course of the runtime of the protocol, we can make the realistic assumption that the factors $\alpha$ and $\beta$ are quite small. This helps us to set a theoretical bound on $\Xi_{stat}$. The variations of mean photon number ($\mu$) and transmissivity ($\eta$) are well within $1 \%$ on an average in our experiment. Therefore, theoretically for setting up the bound we take the values of $\alpha$ and $\beta$ to be around $1 \%$. Additionally, $\Xi_{stat}$ acts as an upper bound for the statistical fluctuations in the number of coincincidences recorded during the course of the protocol.

\subsubsection{Additional figure of merit for security and optimal $\mu$}

We also define an additional figure of merit. The ratio of coincidences to singles. The total number of single detection events can be written as

\begin{equation}
	\label{Eq15:Singles}
	S = Y_1P_1\left(\mu\right) + \frac{1}{2}Y_2P_2\left(\mu\right) + \frac{1}{4}Y_3P_3\left(\mu\right).
\end{equation}

\noindent Using Eq. (\ref{Eq.2:Yields}) and (\ref{Eq.4:Modified_Coincidence}), we arrive at the following expression for the ratio of coincidences to singles, $\zeta$, as follows

\begin{equation}
	\label{Eq.16:Ratio}
	\zeta = \frac{C}{S} = \frac{3\mu^2 + 4\mu}{\mu^2 + 4\mu + 8}.
\end{equation}

\par The above analysis derives parameters that give additional security bounds specific to our protocol. These parameters along with QBER estimation suffice to establish the security of our protocol.

\section{Experimental Implementation and Results}
\begin{figure*}
\centering
  \includegraphics[width=1\textwidth]{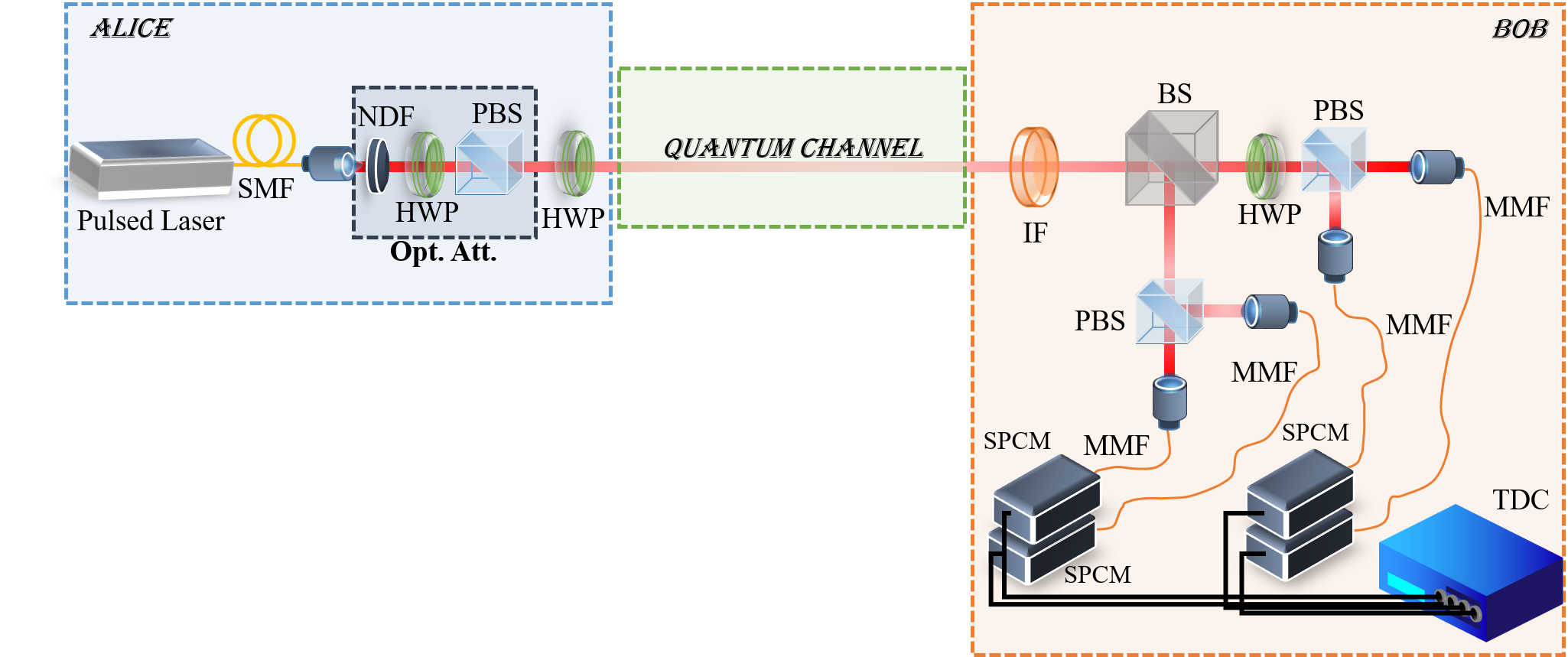}
  \caption{Experimental setup for coincident detection based quantum key distribution protocol. SMF: Single mode fiber; MMF: Multi-mode fiber; NDF: Neutral density filter; HWP: Half-wave plate; PBS: Polarizing beam splitter; BS: 50:50 beam splitter, IF: Interference filter; SPCM: Single photon counting module; TDC: Time to digital converter.}
  \label{fig:setup}
\end{figure*}

We have performed the proof of principle demonstration of our protocol. The details of the experimental setup is shown in Fig. \ref{fig:setup}. 
We have generated weak coherent pulses by using variable optical attenuator at the output of a pulsed laser (Coherent Vitara T (Ti-Sapphire)) with a repetition rate of 80 MHz. After that the encoded state is propagated in free space lossy medium in the laboratory with channel transmissivity estimated at $70 \%$. At Bob's end we have usual polarization based BB84 detection setup: balanced beam splitter (passive random basis selector) with polarizing beam splitter (PBS) on the reflected arm (measurement in $\{$H,V$\}$) and a combination of half wave plate with PBS (measurement in $\{$D, A$\}$) at the transmitted arm. Photons at the output ports of the PBS are detected by fiber coupled avalanche photo diodes (Excelitas SPCM AQRH-14-FC). The avalanche photo diodes are connected to a 8 channel time to digital converter (IDQuantique ID-800) for recording the counts per integration time. It records singles, 2-fold and 3-fold coincidences between various detectors. The coincidence window should be less than or equal to the temporal pulse width of the signal pulse to minimize the probability of a coincidence being recorded between two successive signal pulses or between a signal pulse and any stray pulse. For field applications we can divide our protocol into two categories based on the available channel: I. LOS channel based implementation and II. non-LOS channel based implementation.

\subsection{Direct LOS channel}
Here we propose to use the CD protocol for realistic atmospheric channels where the line of sight between Alice and Bob is under surveillance. This means, Eve's presence can be detected by monitoring the channel by other means and Eve is not allowed to alter the channel transmittance. From application point of view, these assumptions are realistic and give practical security.\\
\begin{figure}[h!]
\centering
\includegraphics[height=7cm,width=11cm]{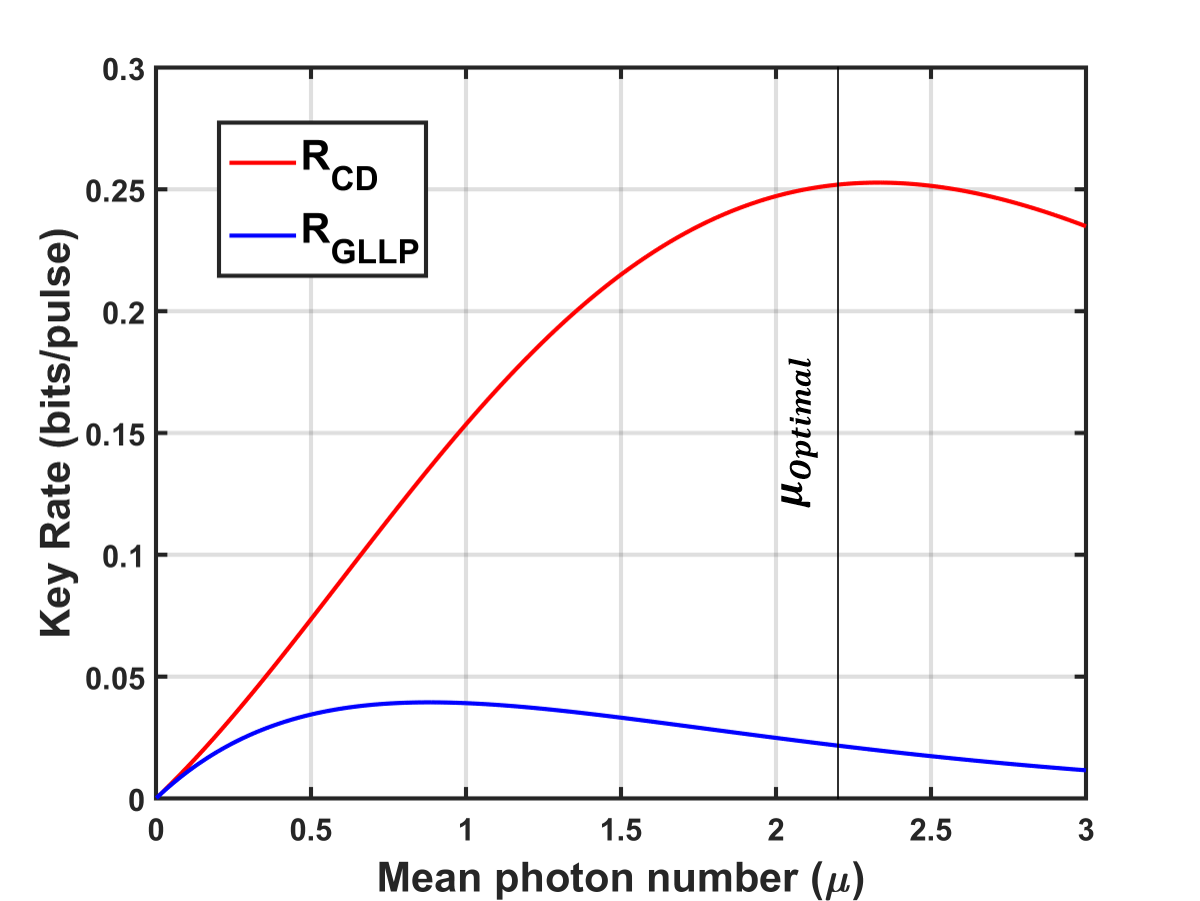}
\caption{Variation of the secret key rate with mean photon number $\mu$ for the GLLP analysis and CD protocol with $\eta=0.70$. As, is evident, the CD protocol has greater tolerance for higher values of $\mu$.}
\label{fig:optimal}
\end{figure}
Here the channel is pre-characterized so the amount of coincidences that Bob will receive is known and is given by Eq.(\ref{Eq.1:Actual_Coincidence}). The key rate for this can then be given by Eq.(\ref{keyrate_final}) and the security comes from observing the figure of merit $\Xi$ defined in Eq.(\ref{Eq.14:FOM_stat}). This results in increase in the optimal $\mu$ for the protocol as given in Fig. \ref{fig:optimal}, which results in increase in the key rate.\\
The channel transmissivity is calculated as the ratio of signals received to signals sent at the detector. This comes out to be $\eta_{t} = 0.70$. $\eta$ can be found from $\eta_{t}$ by dividing it with the efficiencies of detector and the fiber coupler. The yield $Y_n$ and $Q_{\mu}$ can then be calculated by using equations (\ref{yield}) and (\ref{gain}) respectively. We use the calculated value of $\eta$ along with the value of $\mu$ to estimate the number of coincidence events. We list the number of coincidences \emph{C} alongwith $\Xi$ and $\zeta$ in table \ref{tab:coincidence}. It can be seen, the numbers agree within acceptable tolerance with the predicted values from theoretical simulation and as expected, higher values of $\mu$ lead to higher number of coincidences. 
\begin{table}[h!]
\centering
\scriptsize
\caption{List of values for all the security parameters. \emph{C} is the number of coincidences, $\Delta C_{stat}$ is the fluctuation in the number of the recorded coincidences, $\Xi_{stat}$ is the ratio between $\Delta C_{stat}$ and $C$ and $\zeta$ is the ratio between $C$ and the number of detected singles. The numbers in brackets for each of the parameteres are from the theoretical modelling of the protocol for a given channel attenuation. The values of $\alpha$ and $\beta$ are taken to be 0.01 corresponding to a 1 $\%$ variation in the values of $\mu$ and $\eta$ respectively.}
\begin{tabular}{llllll}
\hline \begin{tabular}[c]{@{}l@{}}~~~~$Parameters$~~\end{tabular} 
& \multicolumn{1}{l}{~~~~~~~~~}&{~~~~~}&{~~~~~~~$Values$}&{~~~~~~~~}&{~~~~~~~~} 
\\ 
\hline
\quad $\mu$~~ & \qquad  0.13  & \qquad 0.19 & \qquad 0.22 & \quad 0.32 & \qquad 0.41 \\
\quad $C$~~ & \qquad 3178~(3189)  & \qquad 6249~~(6414) & \qquad 8756~~(8828) & \quad 18367~~(18657) &\qquad 30140~~(30337) \\
\quad $\Delta C_{stat} $~~ & \qquad 53~~(64) & \qquad 69~~(140) & \qquad 85~~(200) & \quad 111~~(250) & \qquad 237~~(340) \\
\quad $\Xi_{stat} $~~ & \qquad 0.016~(0.020)  & \qquad 0.011~~(0.012) & \qquad 0.0097~~(0.023) & \quad 0.0065~~(0.014) & \qquad 0.0079~~(0.11)\\
\quad $ \zeta $~~ & \qquad 0.042~~(0.066)  & \qquad 0.059~~(0.098) & \qquad 0.069~~(0.115) & \quad 0.102~~(0.169) & \qquad 0.128~~(0.218) \\
 \hline 
\end{tabular}
\label{tab:coincidence}
\end{table}
By tracking the number of coincidences, $\Xi$, and $\zeta$, we can monitor the presence of the eavesdropper. If the quantity $\Xi$ is below $\Xi_{stat}$, we can extract keys otherwise the protocol is aborted. Please note that we assume a passive eavesdropper who can only listen in on the communication channel between Alice and Bob and enjoys no control over the channel. In Fig. \ref{fig:keyrate}, we study the secure key rate as a function of the channel length for different values of $\mu$. We see that the secure key rate increases with increasing values of $\mu$ due to increased presence of pulses containing photons. 

\begin{figure}[h!]
\centering
\includegraphics[height=7cm,width=11.5cm]{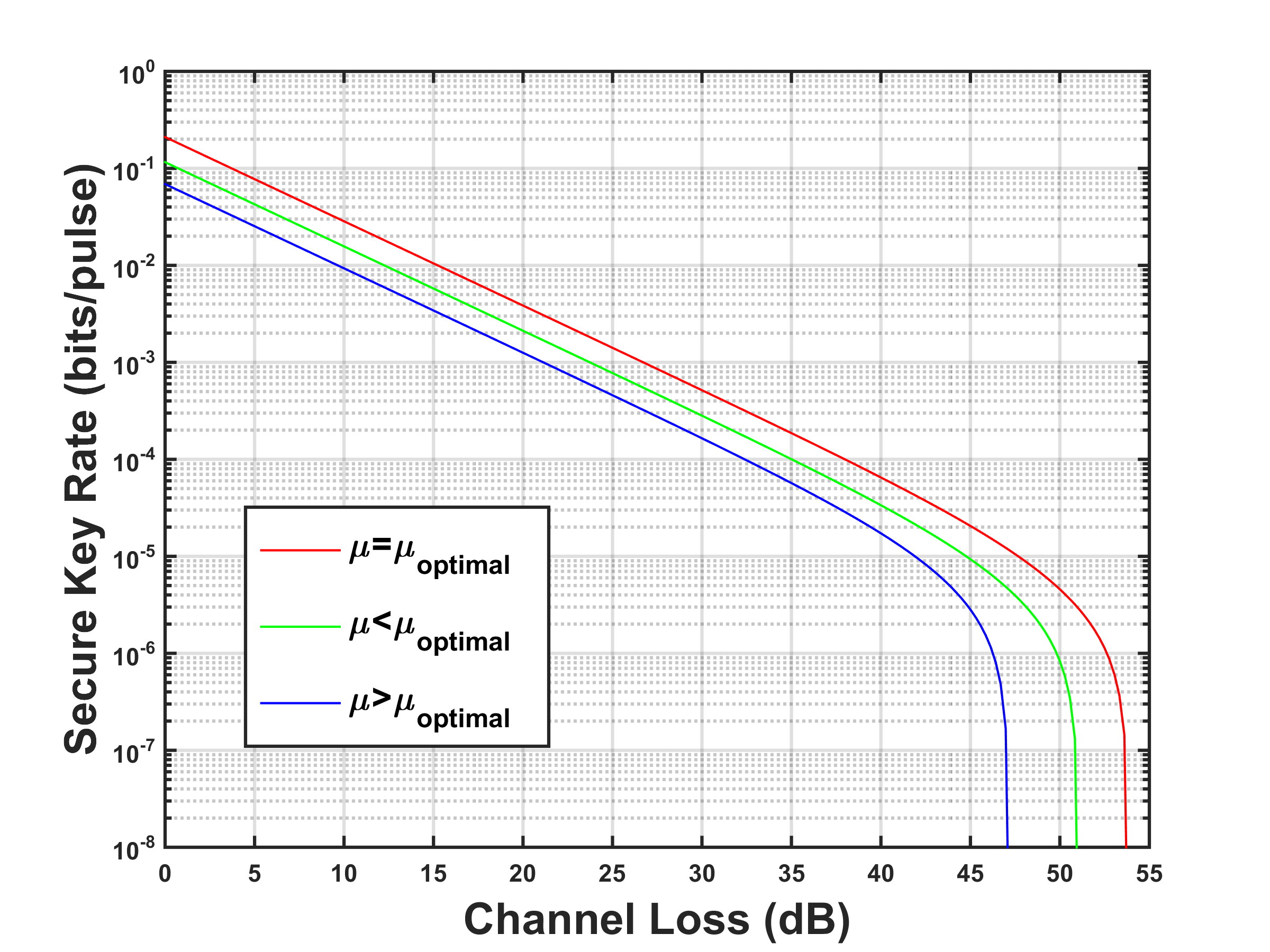}
\caption{Secure key rate as function of the channel length with $\mu$ as a parameter. The value of $\mu_{optimal}$ is obtained from Fig. \ref{fig:optimal} and is equal to 2.2. Two other values of $\mu$ used in the plot are 0.8 ($\mu < \mu_{optimal}$) and 2.9 ($\mu > \mu_{optimal}$)}.
\label{fig:keyrate}
\end{figure}
\begin{figure}[h!]
\centering
\includegraphics[height=7cm,width=11.5cm]{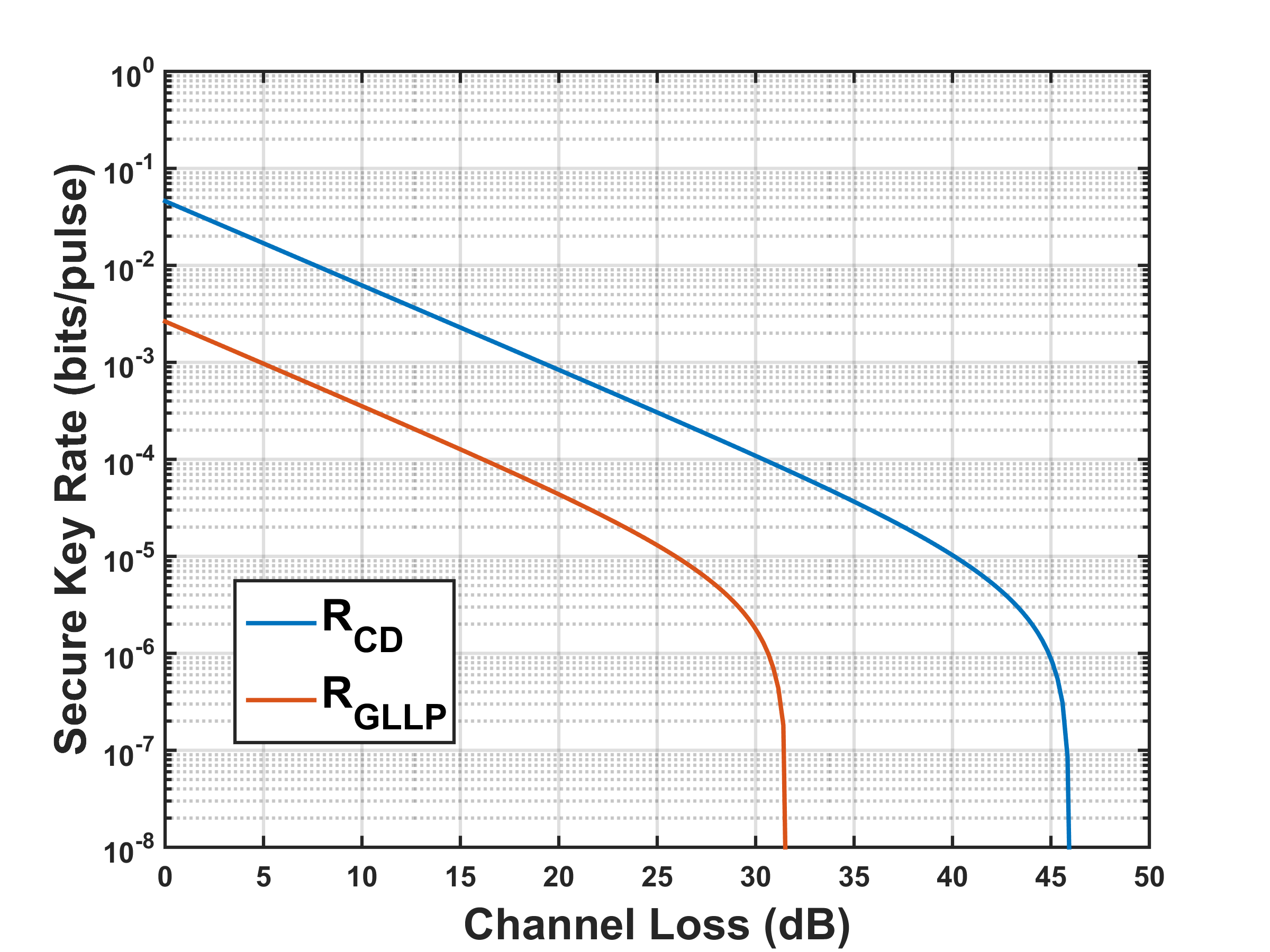}
\caption{Comparison of secure key rates between GLLP analysis and CD protocol for the same set of parameters.}
\label{fig:keyrate_comparison}
\end{figure}

Next, we compare the secure key rates of our protocol with that calculated from \cite{Gottesman:2004:SQK:2011586.2011587} for the same set of parameters, in Fig. \ref{fig:keyrate_comparison}. The results show that we have higher key rate along with increase in the transmission distance. For the given channel and $\mu$ =0.41, we expected a key rate of 0.054 bits per pulse. From the experimental data, we obtained 0.053 $\pm$ 0.004. This matches very well with our theoretical model. For the same set of parameters, using the GLLP analysis, the expected key rate was 0.032 bits per pulse and the experimentally obtained key rate was 0.031 $\pm$ 0.003.

The increase in key rate is due to the fact that some of two and three photon pulses also contribute to the key. In addition, this protocol has greater tolerance to higher values of $\mu$ as shown in Fig. \ref{fig:optimal}. In general, the secure key rate starts decreasing when multiphoton pulses start dominating over single photon pulses. Since coincidence measurements alongwith the security parameters $\Xi$ and $\zeta$ can successfully track and extract key from two-photon and three-photon pulses as well as from all the single photon pulses, this results in a much higher tolerance of mean photon number.
 
\subsection{Non direct LOS channel based implementation}
\par For the case when direct LOS is not available eavesdropping will be easier as regular channel monitoring will be a difficult task. Eve can take the advantage of this and can vary the losses accordingly (tamper the channel) to match with the original channel after extracting photons from each of the multi-photon pulses for gaining information of the key. This can be averted by incorporating the extra pulses with variable intensities randomly in between the signals akin to decoy state protocol. Lack of knowledge about the extra pulses makes Eve randomly attacking both the signal and extra pulses with equal possibility. The ratio is generally 70 (signal) : 30 (extra) so if Eve attacks them equally the relative loss in the detected number of photons for signal and extra pulses will be different. This change can be observed if the timing information is matched for the received signal and extra pulses with the transmitted. Checking the relative loss between the signal and extra pulses (i.e if the loss of signal is not equal to the extra pulses) can reveal the presence of eavesdropper, making the protocol secure. It must be noted that the introduction of extra pulses does not affect the higher key rate achieved through our protocol in comparison to decoy state protocol. The key rate formula will remain the same as it uses the optimal mean photon number ($\mu$) in which the protocol must operate to achieve higher key rate. The presented protocol will require a good spectral and temporal filtering mechanism. For spectral filtering, narrow bandwidth band pass filter has to be used. For accurate temporal filtering, a high speed event timer has to be used with a resolution of picoseconds.

\section{Conclusion}
In this article we have proposed Coincidence Detection based BB84 quantum key distribution protocol with weak coherent pulse under restricted eavesdropping assumption set for LOS channel. We have proposed and derived an analytical expression for the secret key rate taking into account the contribution of pulses with more than one photon in the final key. We argue that by closely monitoring the number of coincidence events arising at the receiver end and matching it with the expected number of coincidences, any attempt at channel tampering can be monitored. We have also presented a security proof in support of our protocol and introduced two figures of merit to verify the security of our protocol. We have shown that this results in a higher key rate over longer distances compared to popular implementations of BB84 based protocols for the same set of parameters. We have also performed a proof-of-principle experiment to verify our predictions. The numbers obtained from the experiment agree quite well with the predicted results. One possible demerit might be the need for accurate characterization of the channel which might limit the implementation scenario to clear line of sight situations. Such a situation is mitigated by introducing extra pulses of variable intensities. Introduction of these pulses provide security like decoy state protocol \cite{lucamarini2015security}. The overall simpler setup is beneficial for free space lossy channel since it can achieve higher key rates over longer distances.

\begin{backmatter}

\bmsection{Funding} This work has been partially funded by DST through QuST program.

\bmsection{Acknowledgments} A.B, A.B, N.J, P.C and R.P.S acknowledge the partial funding support from DST through QuST program/ R. K. acknowledges the support from UK EPSRC through Quantum Technology Hub for Quantum Communications Technology, grant no. EP/T001011/1. The authors would also like to thank Prof. Alexander V. Sergienko for his helpful comments.

\bmsection{Disclosures} Authors declare no conflict of interest

\bmsection{Data Availability} Data underlying the results presented in this paper are not publicly available at this time but may be obtained from the authors upon reasonable request.

\end{backmatter}

%%%%%%%%%%%%%%%%%%%%%%% References %%%%%%%%%%%%%%%%%%%%%%%%%

\end{document}